\documentclass{PoS}

\usepackage{url}

\graphicspath{{figures/}}

\title{Development of a strategy for calibrating the novel SiPM camera of the SST-1M telescope proposed for the Cherenkov Telescope Array}

\ShortTitle{Calibration of the SST-1M camera}
\author{
I.~Al~Samarai\textsuperscript{a},
\speaker{C.~Alispach}\textsuperscript{a},
F.~Cadoux\textsuperscript{a},
V.~Coco\textsuperscript{a},
D.~della Volpe\textsuperscript{a},
Y.~Favre\textsuperscript{a},
M.~Heller\textsuperscript{a},
T.~Montaruli\textsuperscript{a},
A.~Nagai\textsuperscript{a},
T.R.S.~Njoh~Ekoume\textsuperscript{a},
I.~Troyano Pujadas\textsuperscript{a},
E.~Lyard\textsuperscript{g}, 
A.~Neronov\textsuperscript{g},
R.~Walter\textsuperscript{g},
V.~Sliusar\textsuperscript{g},
E.~Mach\textsuperscript{h},
J.~Micha{\l}owski\textsuperscript{h},
J.~Niemiec\textsuperscript{h},
J.~Rafalski\textsuperscript{h},
K.~Skowron\textsuperscript{h},
M.~Stodulska\textsuperscript{h},
M.~Stodulski\textsuperscript{h},
T.~Bulik\textsuperscript{d},
M.~Grudzi{\'n}ska\textsuperscript{d}, 
M.~Jamrozy\textsuperscript{b},
M.~Ostrowski\textsuperscript{b},
{\L}.~Stawarz\textsuperscript{b},
A.~Zagda\'{n}ski\textsuperscript{b}, 
K.~Zi{\c e}tara\textsuperscript{b},
P.~Pa{\'s}ko\textsuperscript{i},
K.~Seweryn\textsuperscript{i},
J.~Borkowski\textsuperscript{c},
R.~Moderski\textsuperscript{c}, 
J.~Kasperek\textsuperscript{k},
P.~Rajda\textsuperscript{k},
D.~Mandat\textsuperscript{m}, 
M.~Pech\textsuperscript{m},
P.~Schovanek\textsuperscript{m},
P.~Travnicek\textsuperscript{m} for the CTA SST-1M Project. \\
\textsuperscript{a}\textit{DPNC - Universit\'e de Gen\`eve,  Switzerland} \\
\textsuperscript{b}\textit{Astronomical Observatory, Jagiellonian University, Poland} \\
\textsuperscript{c}\textit{Nicolaus Copernicus Astronomical Center, Poland} \\
\textsuperscript{d}\textit{Astronomical Observatory, University of Warsaw, Poland} \\
\textsuperscript{g}\textit{ISDC, Observatoire de Gen\`eve, Universit\'e de Gen\`eve, Switzerland} \\
\textsuperscript{h} \textit{Instytut Fizyki J{\c a}drowej im. H. Niewodnicza{\'n}skiego Polskiej Akademii Nauk, Poland} \\
\textsuperscript{i}\textit{Centrum Bada{\'n} Kosmicznych Polskiej Akademii Nauk, Poland} \\
\textsuperscript{k}\textit{AGH University of Science and Technology, Poland} \\
\textsuperscript{m}\textit{Institute of Physics of the Czech Academy of Sciences, Czech Republic.} \\
\textsuperscript{1}\textit{https://www.cta-observatory.org/} \\
E-mail: \email{cyril.alispach@unige.ch}

}

\abstract{CTA will comprise a sub-array of up to 70 small size telescopes (SSTs) at the southern array. The SST-1M project, a 4~m-diameter Davies Cotton telescope with 9 degrees FoV and a 1296 pixels SiPM camera, is designed to meet the requirements of the next generation ground based gamma-ray observatory CTA in the energy range above 3~TeV. Silicon photomultipliers (SiPM) cameras of gamma-ray telescopes can achieve good performance even during high night sky background conditions. Defining a fully automated calibration strategy of SiPM cameras is of great importance for large scale production validation and online calibration. The SST-1M sub-consortium developed a software compatible with CTA pipeline software (CTApipe). 
The calibration of the SST-1M camera is based on the Camera Test Setup (CTS), a set of LED boards mounted in front of the camera. The CTS LEDs are operated in pulsed or continuous mode to emulate signal and night sky background respectively. Continuous and pulsed light data analysis allows us to extract single pixel calibration parameters to be used during CTA operation.}

\FullConference{35th International Cosmic Ray Conference - ICRC217-\\
		10-20 July, 2017\\
		Bexco, Busan, Korea}

\begin{document}

\section{The single mirror small size telescope for the Cherenkov Telescope Array}

Cherenkov telescopes collect Cherenkov light induced by the secondary charged particles of atmospheric gamma-ray showers. Most of the Cherenkov light reaches the ground as a circular pool of about 120 m. By imaging the light produced in the atmosphere the shower direction and energy can be reconstructed. This technique is named Imaging Atmospheric Cherenkov Technique (IACT). \\ 

The Cherenkov Telescope Array \cite{cta} (CTA) will be composed of a southern and a northern array to explore the gamma-ray energies from 20~GeV up to 200~TeV. To explore this wide energy range, CTA relies on a set of large (LST), medium (MST) and small (SST) size telescopes, each covering a specific energy range. The small size telescope array will be located in the southern array with up to 70 SSTs over a large area of approximately 1~km$^2$. It will improve the sensitivity of CTA for high energy gamma-rays  above 3~TeV. The single mirror small size telescope (SST-1M)\cite{sst-1m} is one of the proposed telescope. The telescope optics is based on the Davies-Cotton design\cite{sst-1m-optics} with dish diameter of 4 meters and a wide 9 degrees field of view. The SST-1M camera is currently under commissioning at the University of Geneva and first observations will be in Krakow in summer 2017.  \\

Since CTA will contain up to 70 SSTs it is important to define a robust calibration strategy for the fast validation of large scale production. The calibration methods developed during the commissioning will also be used for the on-site calibration of the photo detection plane (PDP) before and during observation. The data reduction from raw images to calibrated images requires that the baseline, dark count rate, night sky background, gain and crosstalk are fully characterized for optimal gamma-ray reconstruction.

\subsection{SST-1M camera}

The SST-1M Photo Detection Plane (PDP) is composed of 1296 hexagonal silicon photomultipliers (SiPMs) from Hamamatsu (S10943-3739(X)) and of front-end electronics, designed at the University of Geneva, which allows us to observe signal across a large dynamic range; from a single photoelectron (p.e.), up to a thousand photoelectrons (p.e). Compared to photomultiplier tubes, SiPMs can be operated even during moonlight nights as they are not damaged by high night sky background (NSB) level. 

The readout and trigger system is carried out by fully digital electronics system, the DigiCam. It is composed of 3 microcrates, each serving one PDP sector of 432 pixels. One microcrate is the master while two are the slaves. The master microcrate is taking care of the final trigger decision and of combining the data in order to send it to the camera server. The microcrate hardware is composed of 9 digitizer boards with 48 channels. Each channel consists of 12-bit Flash Analog to Digital Converters and 1 trigger board. The three DigiCam microcrates are connected together to ensure data exchange between sector borders for the trigger (see \cite{imen-icrc} for more information on the trigger logic). DigiCam samples data from the PDP at a sampling rate of 250 MHz, 4 ns bin resolution. It features a ring buffer that keeps data in the system while continuing observations. The ring buffers allow dead time free operations (refer to \cite{imen-icrc} for the trigger performances). The readout window of DigiCam can dynamically be configured on request through slow control from 1 to 92 samples. Currently we are running with a window of 50 samples (200 ns) while for calibration purposes we use 92 samples (368 ns). There can be up to 255 events triggered without any time gap between them, in this way also very long events can be recorded.

\subsection{The Camera Test Setup}\label{sec:cts}

\begin{figure}
\begin{center}
\includegraphics[scale=0.6]{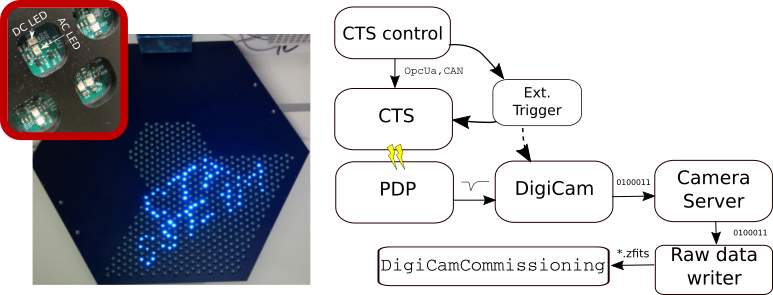}
\end{center}
\caption{Left: Picture of the LED boards of the CTS. Right: Principle of the CTS setup. The CTS control commands the external trigger and the CTS LEDs light level. Triggers are fanned out to the CTS LEDs and DigiCam. Digicam sends the raw data to the camera server that writes it to zfits format. The files are later analyzed by  \texttt{DigicamCommissioning}}\label{fig:cts}
\end{figure}

\begin{figure}
    \centering
    \begin{minipage}{0.45\textwidth}
        \centering
		\includegraphics[scale=0.2]{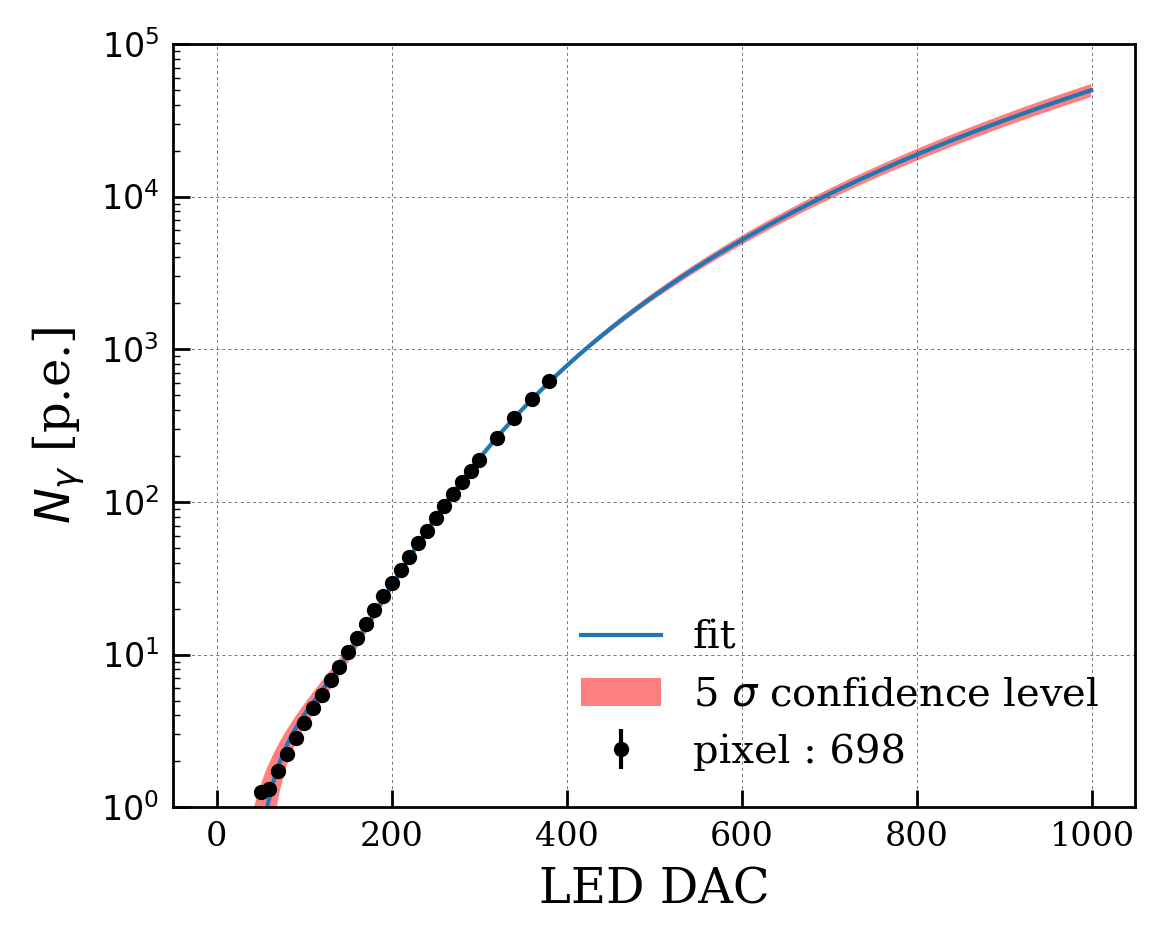}
        \caption{Pulsed LED calibration}\label{fig:ac_calib}
    \end{minipage}\hfill
    \begin{minipage}{0.45\textwidth}
        \centering
		\includegraphics[scale=0.2]{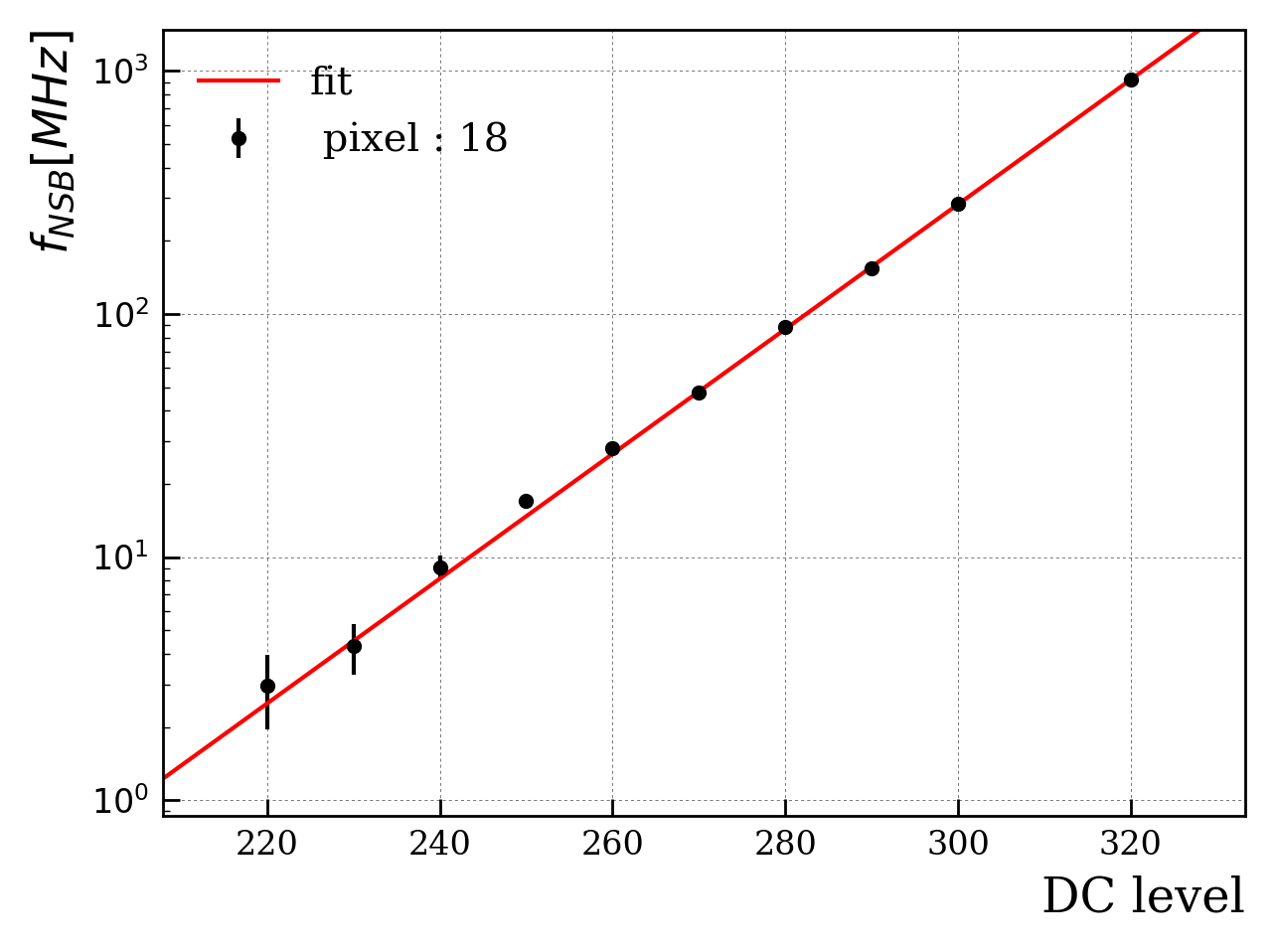}
        \caption{Continuous LED calibration}\label{fig:dc_calib}
    \end{minipage}
\end{figure}

The camera calibration is performed with the Camera Test Setup (CTS), see fig. \ref{fig:cts}. The CTS is an LED board mounted in front of the camera. It is used for single pixel calibration and characterization, and to validate the full camera cabling and behavior. Each pixel faces two LEDs (Broadcom / Avago  ASMT-BB20-NS000), one operated in continuous mode and the other one in pulsed mode. The LEDs have an opening angle of $15^{\circ}$ and a peak wavelength of $470$~nm. The LED voltages are controlled with a DAC. The calibration of the DAC to number of detected photons for pulsed LEDs and DAC to NSB for the continuous LEDs is done on a pixel-by-pixel basis (see fig. \ref{fig:ac_calib} and \ref{fig:dc_calib}) to account for the different SiPM and LED behavior. The LEDs are controlled via a CAN interface by an OPC UA server. The control software of the experimental setup is written in Python. 
In order to collect calibration data, the software controls a pulse generator which sends a trigger signal (exponential rise $\sim 1$~ns and exponential decay $\sim 4$~ns) that is fanned out to the 11 LED boards and to the DigiCam. The channels from trigger signal to the LEDs have been optimized to reduce delay. The DigiCam sends the calibration event to the camera server via optical fiber (10~Gbps link). The raw data is written to a CTA standard format (zfits) by the camera server. The data analysis is then performed within the Python framework \texttt{DigiCamCommissioning}. 

\paragraph{Continuous and pulsed light calibration} 

The pulsed LEDs are calibrated by flashing a sequence of ten thousand light pulses with the CTS for light level varying from 0.5 to 10000 p.e. The corresponding extracted charge form the digital traces are analyzed with \texttt{DigicamCommissioning}. The signal extraction method consist of subtracting the baseline and integrating over 28 ns of signal around the peak position. Each extracted value is filled to a histogram (called Multiple Photoelectron Spectrum). The histograms are then fitted to extract the mean number of detected photoelectrons for each light level. The calibration is processed from single to 1000 photoelectron as CTA requires to characterize the camera in this range.
At higher light level the signal from the SiPM saturates the preamplifier chain. To account for saturation effects a calibrated photodiode was used and showed that the LED calibration curves follow a fourth degree polynomial behavior up to 10000 p.e. \cite{sst-1m}. Therefore the pulsed LED calibration with the CTS is  extrapolated above 1000 p.e. by fitting a fourth degree polynomial (figure \ref{fig:ac_calib}).

Since the SiPM sensors are DC coupled to DigiCam, the amount of continuous light linearly shifts the baseline. The continuous LEDs are thus calibrated by comparing the baseline in dark conditions to the baseline for several light conditions. This feature will us allow to monitor the night sky background during observation. The night sky background induces a continuous current that drops the SiPM overvoltage which as a result lowers the SiPM gain. This effect is further detailed in \cite{gain-drop} and is corrected in the continuous LED calibration (figure \ref{fig:dc_calib}).

\section{A dedicated simulation framework}

A dedicated Monte Carlo simulation of the camera has been developed for the commissioning of the camera. The \texttt{DigicamToy} software 
is a Python framework. It allows to validate the various calibration methods and estimate their systematics. The dark count rate is simulated as a random Poisson process on a readout window and crosstalk using a generalized Poisson distribution \cite{vinogradov}. The electronic noise is emulated as a normal distribution centered around the baseline. Simulations also take into account the impulse response of the SiPMs that are extracted from calibration runs as well as the gain drop. The simulation of the readout consist of a 250 MHz sampling rate and digitization. Accurate comparison of the camera performances can be made. For comparison between data and Monte Carlo refer to \cite{imen-icrc}.

\section{Calibration results using CTS and the toy MC}

\subsection{Single and multiple photoelectron spectra}

\begin{figure}
    \begin{minipage}{0.5\textwidth}
    	\begin{center}
    	\includegraphics[scale=0.18]{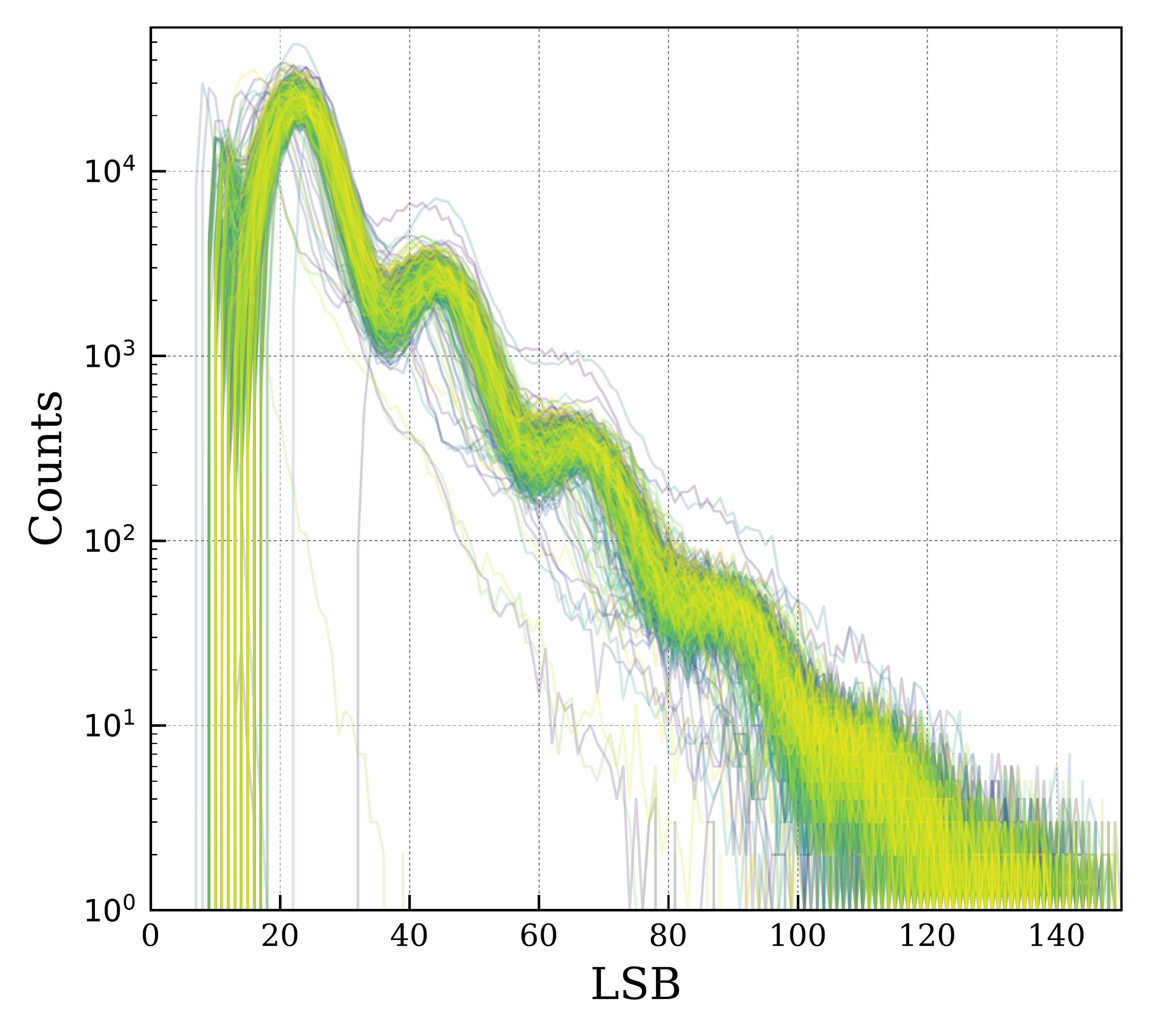}
        \caption{Superposed single photo electron spectra for  one third of the camera }\label{fig:spe}
    	\end{center}
    \end{minipage}
    \begin{minipage}{0.5\textwidth}
		\begin{center}
		 \includegraphics[scale=0.55]{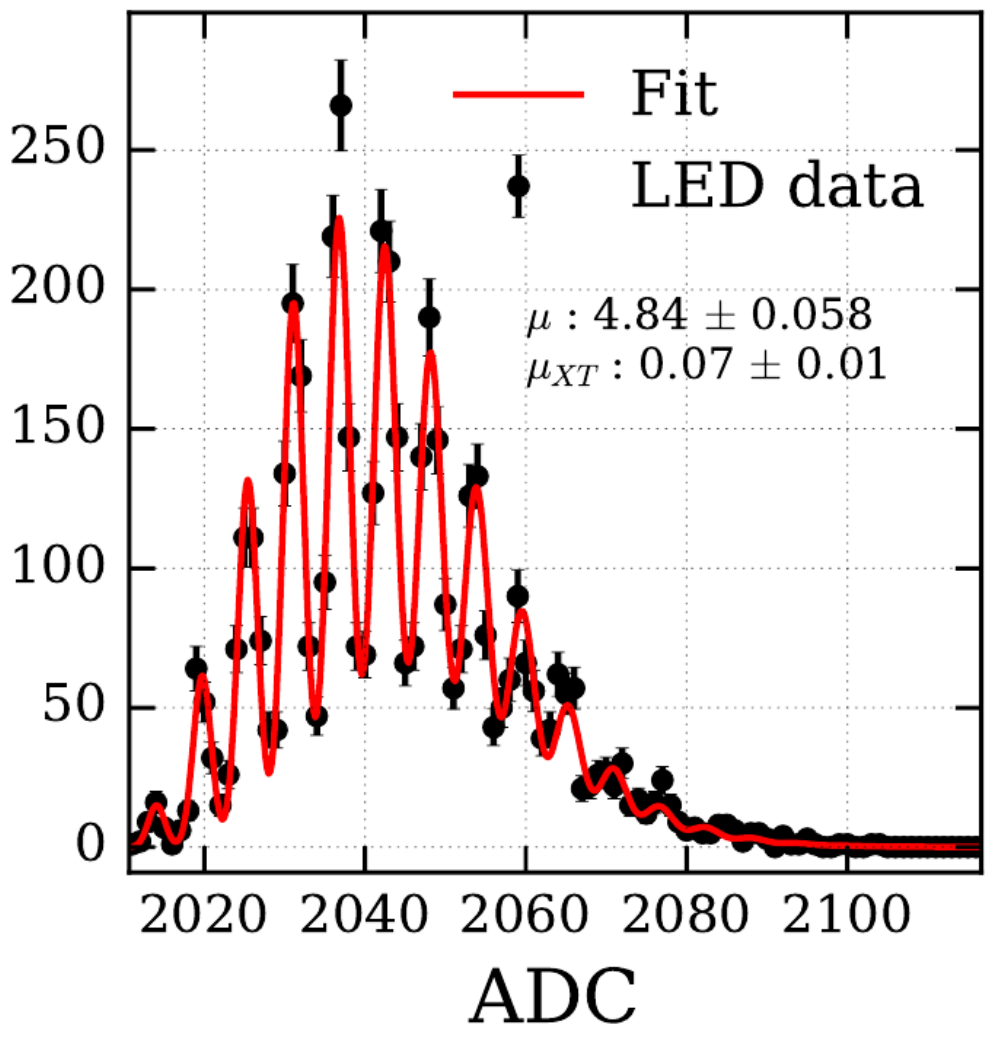}
        \caption{Multiple photo electron spectrum for one pixel of the camera.  The photoelectrons peaks from the pedestal to 13 p.e.  are well distinguished}\label{fig:mpe}
		\end{center}
    \end{minipage}
\end{figure}

The characterization of the camera pixels is based on the study of single and multiple photoelectron spectra (SPE and MPE). We will briefly discuss in this section how both are obtained. The first represents the detector response to a unique photoelectron while the second represents the detector response to a Poisson distributed number of photoelectrons. The SPE is obtained with dark counts events which occur at a rate of $\sim$3~MHz per pixel. About one million events are collected in dark conditions. The photoelectron pulses are detected and the waveform are integrated around the pulse maximum with an integration window of 28~ns. The SPE over one third of the camera is shown in figure \ref{fig:spe}. Further explanation is given section \ref{sec:DCR_XT}. \\

The multiple photoelectron spectrum is obtained from pulsed light measurements. The pulsed LEDs facing each pixel are flashed at 1~kHz in order to collect ten thousand waveforms. They produce an average number of photons distributed according to a Poisson distribution. Due to the small charge spread from micro-cell to micro-cell in the SiPM, the photoelectron peaks are well distinguished.

\subsubsection{Dark count rate and crosstalk}
\label{sec:DCR_XT}

\begin{figure}
    \centering
    \begin{minipage}{0.45\textwidth}
        \centering
		\includegraphics[scale=0.24]{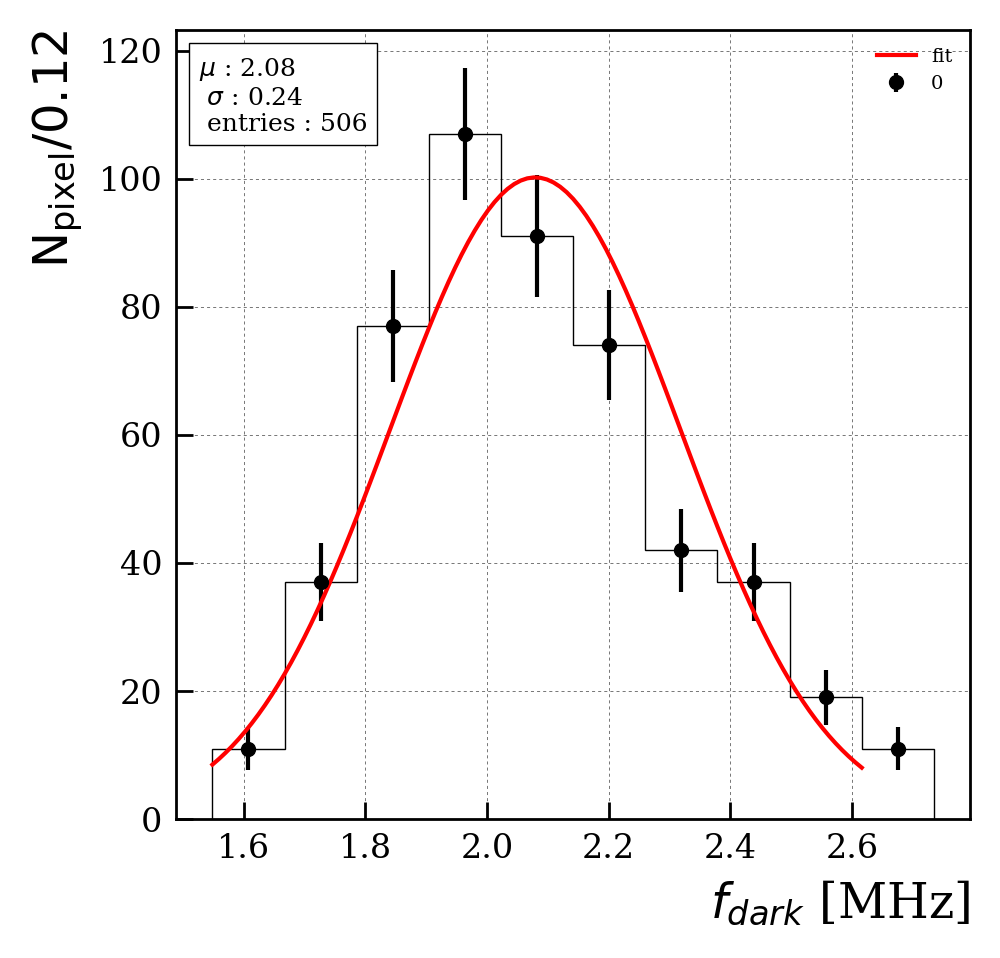}
        \caption{Dark count rate distribution for one third of the camera}\label{fig:dark_count}
    \end{minipage}
    \begin{minipage}{0.45\textwidth}
    \centering
	\includegraphics[scale=0.24]{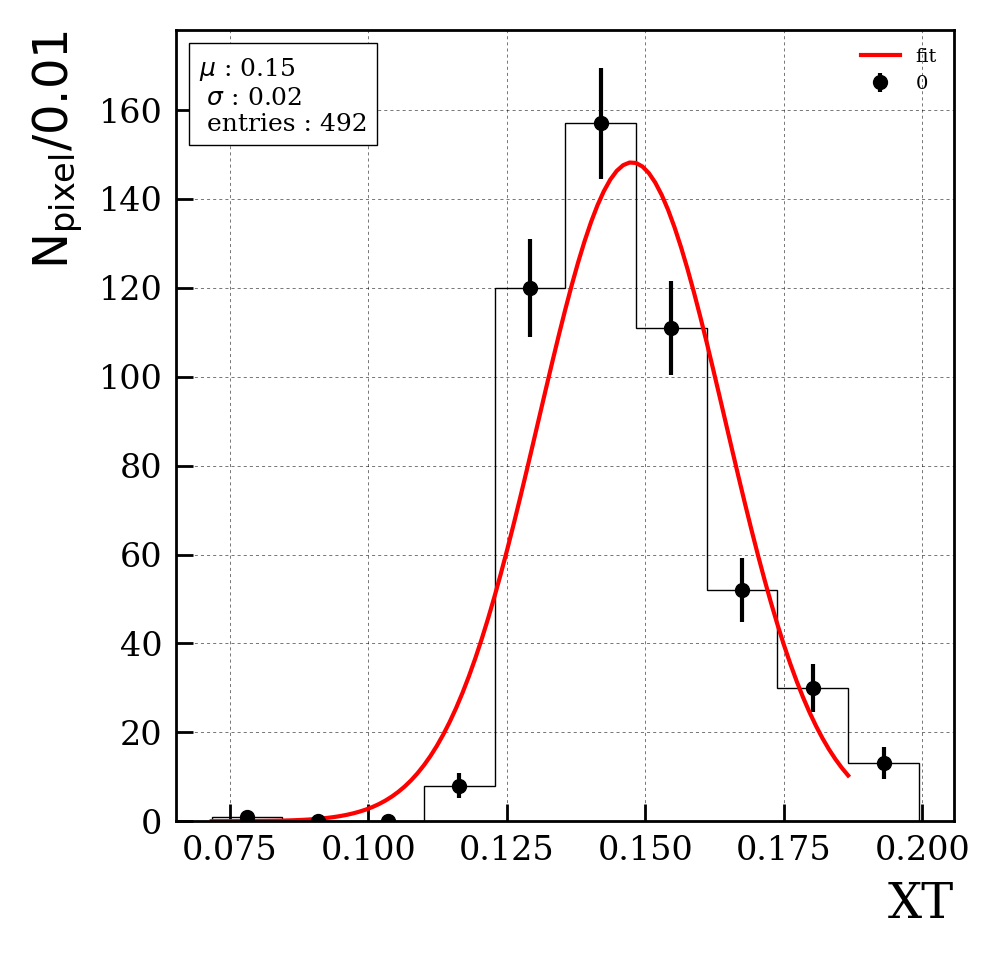}
    \caption{Crosstalk probability distribution for one third of the camera}\label{fig:xt}
    \end{minipage}
\end{figure}

The dark count rate can be obtained with the single photoelectron spectrum by integrating the number of photoelectron pulses recorded divided by the duration of the acquisition. The SPE is obtained with the CTS turned off. The CTS is facing the camera to protect it from external light. Figure \ref{fig:dark_count} gives the distribution of the measured dark count rate over a third of the camera.  \\

After fitting Gaussians below each photopeak, dark count estimated rate is $2.05 \pm 0.2$ MHz per pixel on average. This rate is negligible compared to the expected night sky background which ranges from 40 to 660 MHz per pixel depending on the moon phases. However it is important to determine it for calibration of the CTS and the camera since it biases the measurement in the laboratory conditions. The biases of each calibration parameter are determined with the Monte Carlo. \\

Optical crosstalk occurs when  a secondary avalanche in a micro-cell of the SiPM is triggered by photons produced in the discharge process of a primary cell. As a consequence the amount of charge produced by the SiPM overestimates the amount of detected photons. To correct for this effect, one needs to determine the crosstalk probability, which is usually defined as the ratio between the events with a detected charge greater than 1 p.e. to the total number of events in the SPE spectrum. Figure \ref{fig:xt} shows distribution of the crosstalk probability over a third of the camera with an average 15\% and 2\% standard deviation. However, it is known that this method overestimates the optical cross talk as it neglects the contribution of coincident dark count events. Therefore another description of the optical crosstalk is used. A Poisson branching process as described in \cite{vinogradov} is used. It assumes that each triggered cell can trigger a Poisson number of secondary cells. As a result of the chain process the number of activated cell is distributed according to a Borel distribution if the chain originates from a single cell and according to a Generalized-Poisson if it is from multiple cells. Both descriptions offer a fit distribution for the single and multiple p.e. spectra. The fit function for the MPE is given in eq. \ref{eq:mpe}.

\begin{equation}\label{eq:mpe}
f(x) = \sum_{n=0}^{\infty} P(n|\mu, \mu_{XT})\left[ \frac{1}{\sqrt{2\pi}\sigma_n}e^{-\left(\frac{x-n\cdot G + B}{\sqrt{2}\sigma_n}\right)^2}\right],
\end{equation}

where the first term in the sum, $P(n|\mu, \mu_{XT})$, is given by the generalized Poisson distribution. It represents the probability that $n$ cells are discharged knowing that the average number of photons is $\mu$ and that the crosstalk probability is $\mu_{XT}$. The second term represents the Gaussian smearing through $\sigma_n = \sqrt{\sigma_e^2 + n\sigma_1^2}$, where $\sigma_e$ is induced by electronic noise and $\sigma_1$ the SiPM gain variation from microcell to microcell and the avalanche to avalanche charge spread. The parameter $G$ is the conversion factor from LSB (Least Significant Bit) sum to photoelectron which is more detailed in section \ref{sec:gain} and $B$ represents the baseline i.e. the zero p.e. peak position in the LSB domain. In equation \ref{eq:mpe} all calibration parameters except for dark count rate can be recovered from the fit. Nevertheless there are more specific measurements for certain parameters. For instance, the electronic noise is measured with SiPM being operated below breakdown voltage in order to avoid dark counts. Thus the fitting function is used only to determine $\mu$ and $\mu_{XT}$ (see fig. \ref{fig:mpe}). With the generalized Poisson model of SiPM crosstalk we observe that the crosstalk probability is less than what is obtained using the conventional method (8\% on average for a third of the camera) as expected.

\subsubsection{Gain}\label{sec:gain}

\begin{figure}
\centering
\begin{minipage}{0.45\textwidth}
        \centering
		\includegraphics[scale=0.165]{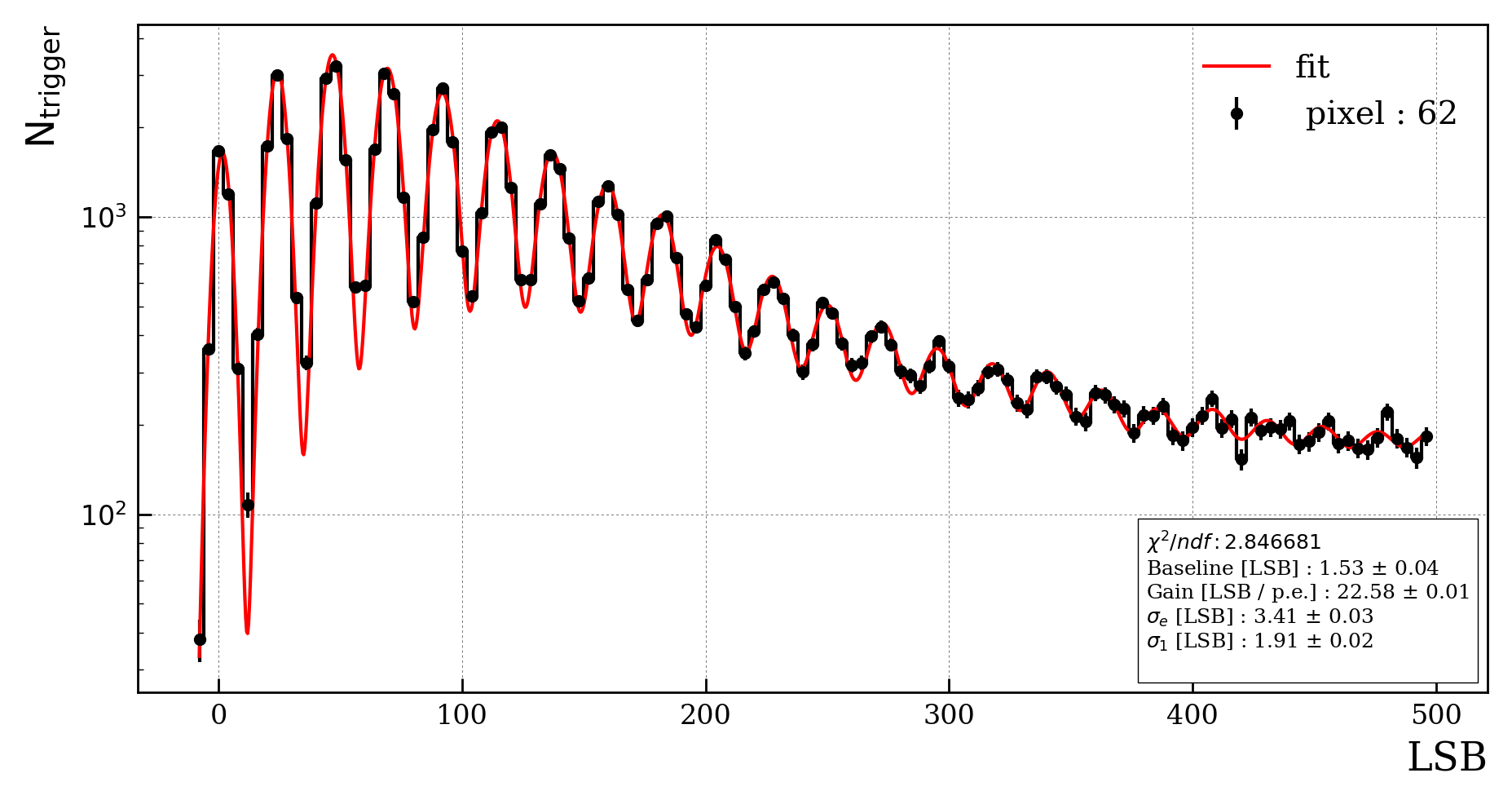}
		\caption{Gain measurements with stacked multiple photoelectron spectra}\label{fig:gain}
\end{minipage}
\hfill
    \begin{minipage}{0.45\textwidth}
    \begin{flushleft}
    \includegraphics[scale=0.22]{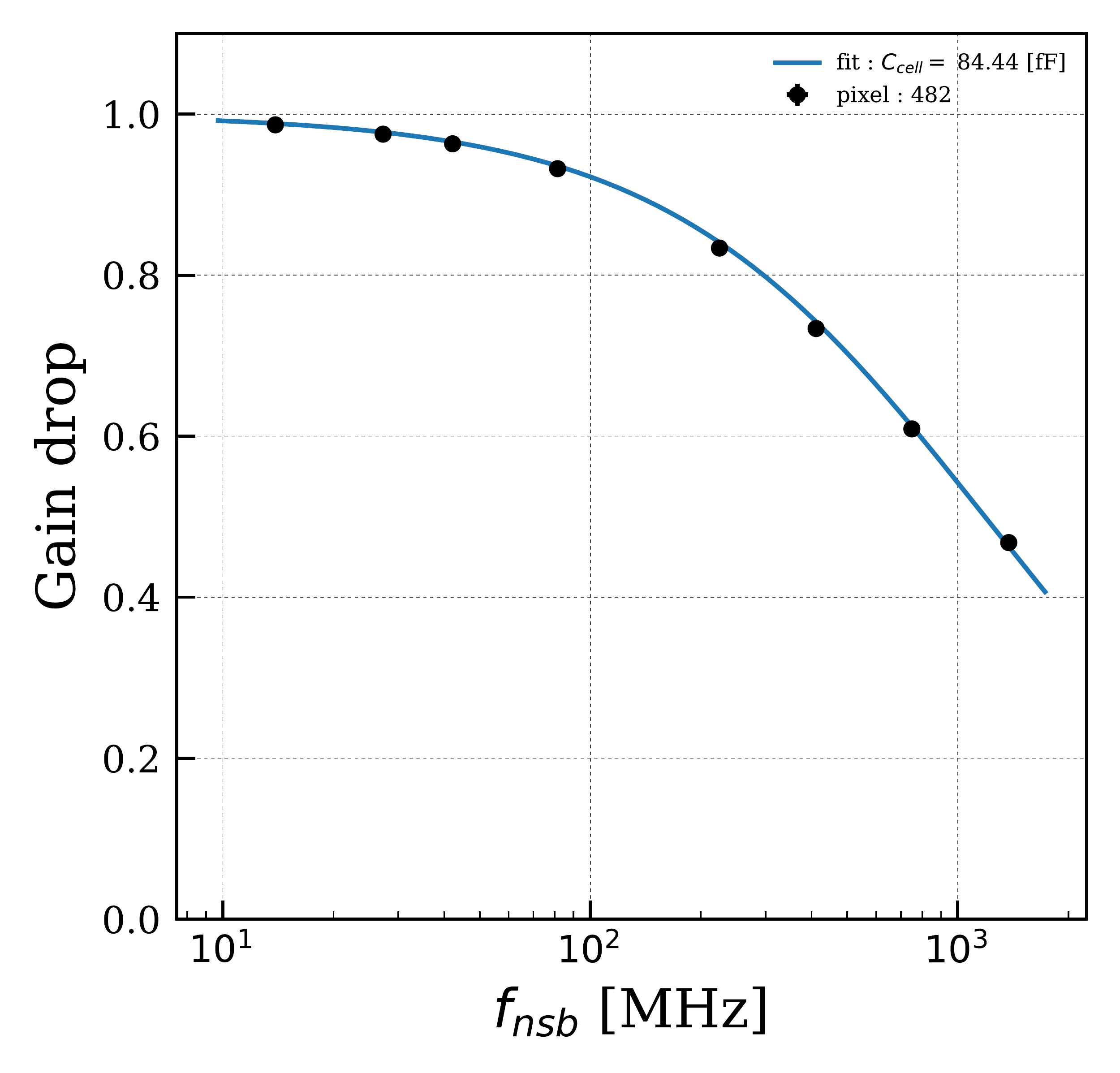}
    \caption{Measured gain drop as function of NSB}\label{fig:gain_drop}
    \end{flushleft}
    \end{minipage}
\end{figure}

The conversion factor form the LSB domain to the p.e. domain, which is often abusively called gain, is measured using a set of combined pulse data from different light levels to build a spectrum. This spectrum (see fig. \ref{fig:gain}) gives a better estimate of the gain since the width of peak to gain ratio is smaller than one (i.e. $\frac{\sigma_n}{G} < 1$) and better statistic is used to constrain the fit.  \\

As mentioned in section \ref{sec:cts} the SiPMs sensors undergo a gain drop effect with increased NSB level. The gain drop effect  tends to damp the baseline shift as the light level increases. More explanation on the gain drop is given in \cite{gain-drop}. To measure the gain drop we compare the waveform amplitude above baseline of fixed pulsed light while increasing the continuous light (see fig. \ref{fig:gain_drop}). Here we observe good agreement between the model and the data. The model parameters value are fitted. A fit from which we obtain a microcell capacitance of 84.44~fF in good agreement with 85~fF provided by Hamamatsu. In order to reduce the gain drop with NSB, the bias resistor ($10$ k$\Omega$) will be replaced by a 2.4 k$\Omega$ resistor. This means that at  660 MHz (half moon night) the gain drop  will be at $88$ \% compared to the current value of $64$ \%.

\paragraph{Systematics study with Monte-Carlo}

The development of a Monte Carlo tool for the SST-1M camera was motivated by the study of systematics of each calibration parameters induced by the signal reconstruction method or by effects that are not taken into account in the fitting function such as dark count rate. The Monte Carlo is interfaced with the CTS such that for each request of calibration data a similar Monte Carlo dataset can be produced. The Monte Carlo dataset is processed by the same analysis pipeline (\texttt{DigicamCommissioning}). For instance, a study of the systematics on the gain determination by introducing dark count and cross talk is summarized in table \ref{tab:bias}. The results show that dark count and crosstalk do not affect much the gain measurements.  However, a larger impact is seen on the baseline, the electronic noise and the SiPM gain smearing. The impact is understood as the dark counts add a non symmetric contribution to the Gaussian nature of the p.e. peaks.

\begin{table}
\centering
\resizebox{0.8\columnwidth}{!}{%
\begin{tabular}{l||c|c|c|c}
~ & True & $\mu_{XT} = 0$ [p.e.] & $\mu_{XT} = 0$ [p.e.] & $\mu_{XT} = 0.06$ [p.e.] \\ 
~ & ~ & $f_{dark} = 0$ [MHz] & $f_{dark} = 3$ [MHz] & $f_{dark} = 3$ [MHz] \\ 
\hline \hline
gain [LSB/p.e.] & $5.6$ & $5.6 \pm 0.00$ & $5.6 \pm 0.01$ & $5.6 \pm 0.00$ \\
baseline [LSB] & 2010 & $2010 \pm 0.02$ & $2010.06 \pm 0.01$ & $2010.07 \pm 0.01$ \\
$\sigma_e$ [LSB] & 0.86 & $0.86 \pm 0.01$ & $0.90 \pm 0.02$ & $0.9 \pm 0.01$ \\
$\sigma_1$ [LSB] & 0.48 & $0.48 \pm 0.00$ & $0.49 \pm 0.01$ & $0.49 \pm 0.06$ \\
$\chi^2$/ndf & & $1 \pm 0.17$ & $ 1.16 \pm 0.19$ & $1.74 \pm 0.22$\\
\hline
gain pull [LSB/p.e.] & & $-0.17 \pm 1$ & $-0.17 \pm 0.97$ & $0.65 \pm 1.02$\\
baseline pull [LSB] & & $-0.17 \pm 1.02$ & $-2.63 \pm 1.07$ & $-7.11 \pm 1.02$ \\
$\sigma_e$ pull [LSB] & & $-0.11 \pm 1.07$ & $-2.27 \pm 1.06$ & $-5.25 \pm 1.12$ \\
$\sigma_1$ pull [LSB] & & $-0.05 \pm 0.99$ & $-1.20 \pm 1.07$ & $-2.92 \pm 1.10$ \\
\end{tabular}}
\caption{Results of a study of systematics on gain measurements introducing dark count and crosstalk. The true values are given in the first column while the measured are in the following columns. The pulls represent the difference between the true and  measured value in terms of 1$\sigma$ error bar.}\label{tab:bias}
\end{table}

\section{Conclusion}

In this contribution we defined and explained the calibration strategy. The CTS and its components were presented. Calibration results from the CTS show that the SST-1M sub-consortium has established robust and reliable calibration protocols for the foreseen implementation of the SSTs in the CTA southern array. In the future the calibration will be repeated, using a flasher developed at LUPM, during the operation of the telescope.

\section*{Acknowledgements}

This work was conducted in the context of the CTA Consortium SST-1M Project. We gratefully acknowledge support from the University of Geneva, the Swiss National Foundation, the Ernest Boninchi Foundation and from the agencies and organizations listed here : \url{http://www.cta-observatory.org/consortium_acknowledgments}. In particular we are grateful for support from the NCN grant DEC-2011/01/M/ST9/01891 and the MNiSW grant 498/1/FNiTP/FNiTP/2010 in Poland. The authors gratefully acknowledge the support by the projects LE13012 and LG14019 of the Ministry of Education, Youth and Sports of the Czech Republic. This paper has gone through internal review by the CTA Consortium.


\begin{thebibliography}{99}
\bibitem{cta} CTA Consortium, Astropart.Phys. 43 (2013) 3-18
\bibitem{sst-1m} M. Heller et al, Eur. Phys. J. C (2017) 77-47
\bibitem{sst-1m-optics} K.Seweryn for the SST-1M project, Proceedings of Science, The 34th International Cosmic Ray Conference, 2015
\bibitem{imen-icrc} I. Al Samarai for the SST-1M project,  \pos{PoS(ICRC2017)758} (these proceedings).
\bibitem{vinogradov} S. Vinogradov, Analytical models of probability distribution and excess noise factor of solid state photomultiplier signals with crosstalk,  arXiv:1109.2014
\bibitem{gain-drop} J. A. Aguilar et al., Proc. SPIE 9915, High Energy, Optical, and Infrared Detectors for Astronomy VII, 99152T (27 July 2016)
\end{thebibliography}
\end{document}